\documentclass[preprint,12pt]{elsarticle}

\usepackage{graphicx}

\usepackage{amssymb}

\journal{Journal of Computational and Applied Mathematics}

\begin{document}

\begin{frontmatter}



\title{A numerical method for imaging of biological microstructures by VHF waves}


\author[a]{Guido Ala}
\author[b]{Pietro Cassar\`a}
\author[c]{Elisa Francomano\fnref{label1}}
\fntext[label1]{Corresponding author: Elisa Francomano, \textit{E-mail address}:elisa.francomano@unipa.it }

\author[a]{Salvatore Ganci}
\address[a]{Universit\`a degli Studi di Palermo, DEIM, Palermo, Italy}
\address[b]{CNR, Istituto di Scienza e Tecnologie dell'Innovazione,Pisa,Italy}
\address[c]{Universit\`a degli Studi di Palermo, DICGIM, Palermo, Italy}

\begin{abstract}
Imaging techniques give a fundamental support to
medical diagnostics during the pathology discovery as well as
for the characterization of biological structures. The imaging
methods involve electromagnetic waves in a frequency range that
spans from some Hz to GHz and over. Most of these methods
involve scanning of wide human body areas even if only small
areas need to be analyzed. In this paper, a numerical method
to evaluate the shape of micro-structures for application in the
medical field, with a very low invasiveness for the human body,
is proposed. A flexible thin-wire antenna radiates the VHF waves
and then, by measuring the spatial magnetic field distribution it is
possible to reconstruct the micro-structure’s image by estimating
the location of the antenna against a sensors panel. The typical
inverse problem described above is solved numerically, and first
simulation results are presented in order to show the validity and
the robustness of the proposed approach.
\end{abstract}


\end{frontmatter}


\section{Introduction}

The support given by the imaging to medical diagnostics
is fundamental during the pathology discovery as well as
for biochemical characterization of biological structures \cite{bib_1}-\cite{bib_7}.
The imaging methods involve electromagnetic waves in a
frequency range that spans from some Hz to GHz and over.
Hence, the understanding of biological structures response to
the electromagnetic field is fundamental. The investigation
of the interaction between healthy or pathological biological
tissues and electromagnetic waves is a hot topic yet \cite{bib_8}. 
In fact,understanding how an electromagnetic wave interacts with the
human body becomes increasingly important if one considers
that the most of imaging methods involve scanning of wide
areas of the human body, even if only small areas need to
be analyzed. In this way, although a wide scanning allows to
acquire a big amount of data by single exposition, also areas
of the body not interested in the diagnostics are exposed to
the waves, so increasing the invasiveness for the patient. For
this reason, new imaging systems able to analyze only the area
under diagnostics (confined scanning), are becoming more and
more investigated.
A big input to micro-imaging systems has been given by the
micro-electronic technology, which allows the development
of systems for the scanning of confined small areas. In fact,
emitters and receivers (sensors) smaller and smaller can be
made, so that the final size of the imaging systems as well as
the wave’s spot become very small.
Generally, the acquired data need to be appropriately elaborated
extracting the imaging information by means of inverse
optimization algorithms \cite{bib_3},\cite{bib_4},\cite{bib_7}. Through these algorithms
high quality information can be extracted from the data
acquired by the sensors, even if the quality of the sensor’s
signal is low, because of their small area which may determine
a significant amount of noise.
In this paper a method to elaborate the shape of microstructures
for application in the medical field is proposed.
The method works with low power waves in the Very High
Frequency (VHF) range, in order to achieve a low invasiveness
for the human body.
The method uses a system endowed with: a microtransmitter
to emit a magnetic field, a sensors panel to acquire the spatial
distribution of the magnetic field and an elaboration logic
to acquire and elaborate the sensor’s signals. The microtransmitter
radiates the VHF waves by means of a microantenna,
which is able to take the shape of the target structure.
If the micro-antenna is assumed to be a sequence of thin-wire
interconnected short dipoles, then it is possible to reconstruct
the micro-structure’s image by measuring the spatial distribution
of the magnetic field. In fact, the shape reconstruction is
possible by estimating the location of thin-wire antenna against
the sensors panel.
The recognition problem of the thin-wire antenna’s location
by magnetic field can be addressed as an inverse problem. The
thin-wire antenna is supposed to be a sequence of linear segments:
given a model for the characterization of the magnetic
field at a point in space (forward problem) and given a set of
measurements of the magnetic field amplitude, it is possible
to solve the inverse problem in terms of the distance of the
antenna from the sensors panel.
In this paper, preliminary results about the previous basic
idea, are reported concerning two simulated scenarios.
Namely, the spatial distribution of the emitted magnetic field
is simulated through a numerical model based on the method
of moments (MoM), where the first kind Fredhlom’s integral
equation is solved by the point matching procedure \cite{bib_9}-\cite{bib_15}.
The spatial distribution of the magnetic field is evaluated on
an area equivalent to the area of the sensors panel. These
magnetic field values are used as the measured input for the
inverse problem. The Levenberg-Marquart algorithm \cite{bib_16}, \cite{bib_17}
is considered in solving the inverse problem by means of the minimization of the Euclidean distance between the measured
field and the field generated by a given configuration of thinwire
piecewise antenna. The numerical procedure involved by
the algorithm rounds the entries of the Hessian matrix, step
by step, and the location of the antenna’s segments can be
estimated with high precision, so obtaining the image of the
shape taken by the antenna.
The paper is structured as follows. In section 2 the mathematical
framework about the proposed method is presented.
In section 3 some numerical results about reconstruction
examples are discussed, then a conclusion about the method
results and ongoing work complete the paper.
\section{Numerical approach}
\label{}
The shape of a biological thin micro-structure can be estimated
by means of an appropriate embedded emitting antenna
and by measuring the spatial distribution of the opportune
radiated electromagnetic field component. In particular, the
magnetic field may be the best choice together with the selection
of the appropriate work frequency range. The proposed
approach is a typical inverse problem: the measured spatial
distribution of the magnetic field emitted by a VHF thinwire
antenna (i.e. with radius less than one hundredth of the
maximum work wavelength) is the input, and the source shape
reconstruction is the final target.
The antenna is assumed to be a piecewise linear structure,
i.e. a sequence of $\textit{N}_{\textit{b}}$ branches with uniform radius  
$\textit{r}_{\textit{cond}} $ and conductivity \begin{flushright}
\end{flushright}$\sigma_{omega} $ . The antenna is fed at the point $P_{0}(x_{0},y_{0},z_{0})$ by a sinusoidal current source with known
frequency$ \textit{f}$ and amplitude $I_{source}$. The spatial distribution of the magnetic field is detected by a sensors panel with $\textit{M}$ sensors distributed on its surface. As a first approximation,
the surrounding medium is supposed to be homogeneous,
isotropic, with conductivity $\sigma$, electric permittivity $\epsilon$ and
magnetic permeability $\mu$.
Under these assumptions, the shape reconstruction involves the minimization of an objective function shown in equation (1), by using an iterative procedure. Roughly speaking, step by step the iterative procedure looks for a set of segments coordinates that minimize the difference between the computed
and measured magnetic field at sensor points.
\begin{equation}
\label{minimo}
min_{p}\left\lbrace 
||H_{comp}(p)-H_{means}(p)||_{2} 
\right\rbrace 
\end{equation}

In equation (\ref{minimo}) $ p = [(x_{1}, y_{1}, z_{1})...(x_{N_{b}}, y_{N_{b}}, z_{N_{b}})]^{T} $  is the vector of the cartesian coordinates of the antenna segments ends to be estimated, $H_{comp}$ is the vector of the magnetic field values at sensors points computed by assuming that the
segments ends are located at p, and $H_{meas}$ is the vector with
the measured magnetic field values. The values of $H_{comp}$ are
calculated by the forward solver.
The forward solver computes the amplitude of the magnetic
field at a set of points, for a given signal source and antenna’s
characteristics. More precisely, the current distribution along
the antenna is evaluated by solving an appropriate integral
equation in frequency domain, derived from Maxwell’s equations.
Then, by using the relation that link the currents to
the magnetic field, this latter can be calculated at the sensors
points. The equations of the forward model is solved numerically.
For the problem addressed in this case, the moments
method (MoM) via point-matching procedure in the frequency
domain, is performed . Note that, by this method the integral
equations involved include the boundary conditions, so that
only the emitter have to be discretized but not the problem’s
domain. This means that each antenna’s branch needs to be
split in a finite number of linear segments \cite{bib_9}-\cite{bib_14}.
The problem formulation leads to a first kind Fredholm’s
integral equation. In fact, by expressing the electric field  $\vec{E}$ with the retarded magnetic vector potential $\vec{A}$, the source frequency $(\omega=2\pi f)$ and the scalar potential $\phi$, as shown in equation (2):
\begin{equation}
\vec{E}=-(\nabla \phi+ j \omega \vec{A} )
\end{equation}
by introducing the boundary conditions by means of the
per-unit-length surface impedance $\dot{z_{s}}$, the following equation holds:

\begin{eqnarray}
\vec{\textbf{u}}\cdot \vec{E}=\dot{z_{s}}I_{s}
\end{eqnarray}
\begin{center}
$-(\nabla \phi+ j \omega \vec{A} )_{tg}=z_{s}I_{s}$
\end{center}
in which \textit{tg} indicates the component tangential to the wire surface, $I_{s}$ is the longitudinal current flowing into the conductor, concentrated on its axis $( \textbf{u}'$ as the unit vector) because of the thin-wire assumption \cite{bib_9}, \cite{bib_12},  $\textbf{u}$ is the unit vector tangential to the conductor’s surface, as shown in figure 1.

\begin{figure}[ht]
\label{fig:fig1}
\center
\includegraphics[width=6cm]{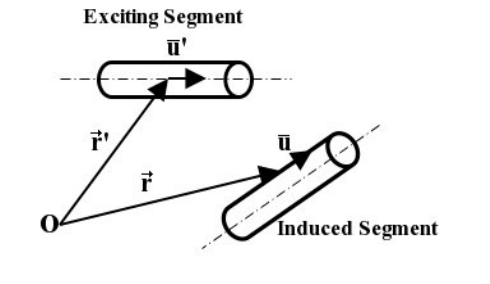}
\caption{General thin-wire geometric references.}
\end{figure}

By using the relation between the magnetic vector potential $\vec{A}$ and the conduction current density, and the relation between the scalar electric potential $\phi$ and the free electric charges density, the well known electric field integral equation (EFIE) is obtained. Then, by integrating the EFIE along the surface of the conductor, the following modified EFIE is obtained \cite{bib_9}, \cite{bib_12}:

\begin{eqnarray}
\begin{array}{lll}
\label{EFIEm}
j\omega\int_{L}\vec{\textbf{u}}\cdot \int_{\Omega}\mu I_{s}(l')
g(\vec{r},\vec{r'}) dl'dl \\
\\
+\int_{L}\frac{\partial}{\partial l_{tg}}\frac{1}{j\omega\dot{\epsilon}}\int_{\Omega}
\frac{dI_{s}(l')}{dl'}g(\vec{r},\vec{r'}) dl'dl=\\
\\
 \dot{z_{s}}\int_{L} I_{s}(\vec{r}) dl
\end{array}
\end{eqnarray}
Equation (\ref{EFIEm}) is a general relation that depends only on thelongitudinal current and on geometrical quantities of the
conductors constituting the thin-wire structures to be analyzed.
In fact, $\Omega$ is the length of the exciting conductor,
\textit{L} is the length of the induced conductor, $\vec{r} $ and $ \vec{r'}$ are space position vectors of the observation and source points, respectively, $g(\vec{r},\vec{r'})=\frac{e^{-k|\textbf{r}-\textbf{r'}|}}{4\pi|\textbf{r}-\textbf{r'}|}$ is the Green’s function in
an unbounded region. The quantity $\dot{\epsilon}=\epsilon+j\omega\sigma$ take into
account the complex medium permittivity and $\dot{k}=\sqrt{-\omega^{2}\mu\dot{\epsilon}} $
is the wave number.
As already underlined, the forward problem, represented by
equation (\ref{EFIEm}), is numerically solved by splitting the thin-wire antenna into a finite number of linear segments of length $\Delta$.
In scientific literature numerous results \cite{bib_9}, \cite{bib_12} show that
an acceptable accuracy can be obtained for the solution of
equation (4), by assuming $\Delta\leq 0.05\lambda$  and a linear distribution
of current along each segment. In this way, a linear system of
order $n+N_{b}$ is obtained with \textit{n} as the number of segments.
Once the currents are computed, the magnetic field components
in the surrounding medium are given by the dipole theory, by superposing the effects of all segments  \cite{bib_9}, \cite{bib_12}.
As discussed above, once the direct solver is obtained, the
inverse problem can be then approached. In figure 2 the
blocks diagram of the algorithm used to solve the inverse
problem is shown.

\begin{figure}[ht]
\label{fig:fig2}
\center
\includegraphics[width=8 cm]{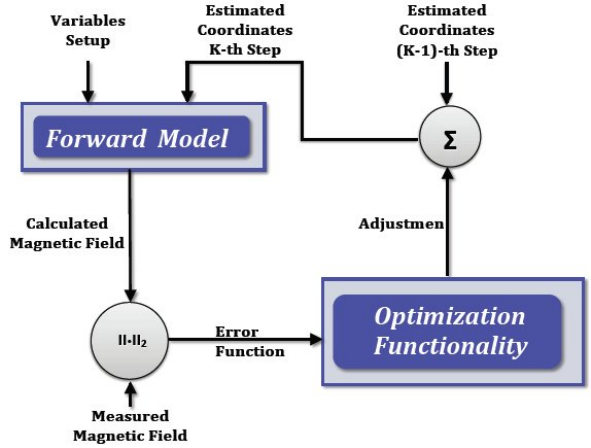}
\caption{Blocks diagram of the inverse problem algorithm.}
\end{figure}

The unknown end-points segments coordinates s are obtained
through an iterative procedure which, at each step,
computes the correction factors needed to obtain the new set
of coordinates. The correction factors are evaluated by the
optimization functionality, on the basis of the difference in \textit{2}-norm between the measured fields and the fields computed by the forward solver. Since the problem is strictly nonlinear, the solver of the inverse problem was based on the Levenberg-Marquardt (LM) algorithm [15].
In brief, the LM algorithm starts from an initial set of
reasonable end-points segments coordinates $s^{(0)}$: this set is
then updated at each step \textit{k} by adding the solutions  $\Delta s^{(k)}$ of the linear system shown in equation (\ref{LM}) 

\begin{equation}
\label{LM}
\left(\tilde{\textbf{B}}^{(k)}+\beta^{(k)}\textit{diag}
[\tilde{\textbf{B}}^{(k)}]\right)\Delta \textbf{s}^{(k)}=
-\textbf{G}^{(k)^{T}}\textit{F}(\textbf{s}^{(k)})
\end{equation}

In equation (\ref{LM}) the matrix $\tilde{\textbf{B}}^{(k)}$ is an approximation of the Hessian matrix of $\textit{F}$ at step $\textit{k}$ and is equal to $\textbf{G}^{(k)^{T}}$ $\textbf{G}^{(k)}$, with $\textbf{G}^{(k)}$ as the gradient of the error function $\textit{F}$ at step $\textit{k}$, while $\beta^{(k)}$ is an adaptive parameter. Once \textbf{$\Delta$} $\textbf{s}^{(k)}$ is known, the new coordinates are calculated by the following equation:
\begin{equation}
\label{correction}
\textbf{s}^{(k+1)}=\textbf{s}^{(k)}+\Delta \textbf{s}^{(k)}
\end{equation}

\section{Numerical results}
In order to validate the capabilities of the proposed method,
some numerical experiments concerning the position estimation
estimation of the antenna’s branches are carried out.
The thin-wire antenna is assumed to be made by a nickel titanium
thin wire 0.1 mm in radius, 2.1 cm in length with
an electrical conductivity of $1.1\cdot{10^{6}}$ S/m. In this way the
$\textit{thin wire}$ assumption is verified. The nickel-titanium alloy is
selected because it is the most diffused alloy in the biomedical
applications. It has to be underlined that this type of antenna
can be effectively realized in practice and some of the human
cavities may be investigated. Moreover, the number of
branches for the antenna is set to $N_{b} = 4$, with shape shown in figure 3 and figure 4 by the red dashed line with squared markers.
The antenna is fed by a sinusoidal current source with
frequency equal to 100 MHz, and amplitude spacing in the
following set of values: 0.1 mA, 1 mA, 5 mA, 25 mA and
50 mA. It has to be underlined that all these current values as well as the selected frequency can be well tolerated by the human tissues. \\The ground surface of the antenna lies on the $\textit{Y} -\textit{Z}$ plane as showed in figure 3 and figure 4, and the feed point is placed at a distance of 2 mm from the ground plane.
\begin{figure}[ht]
\label{fig:fig3}
\center
\includegraphics[width=8cm]{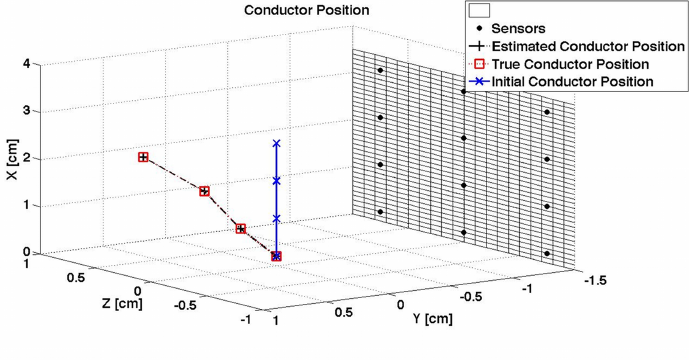}
\caption{ Conductor’s position estimated by flat sensor panel: the initial position is that a of straight thin-wire antenna 2.1 cm length, perpendicular to x = 0 plane.}
\end{figure}
Two sets of 12 sensors are considered. The sensors are
assumed to be sensitive to the component along the $\textit{Z}$ axis of the magnetic field and placed on a flat surface in a first case study. \\A semi-cylindrical surface sensors panel is then also considered (figure 4). The flat sensor panel is 3.5 cm height
and 1.5 cm width, while the semi-cylindrical one has 1.5 cm radius and 3.5 cm height.
\begin{figure}[ht]
\label{fig:fig4}
\center
\includegraphics[width=8cm]{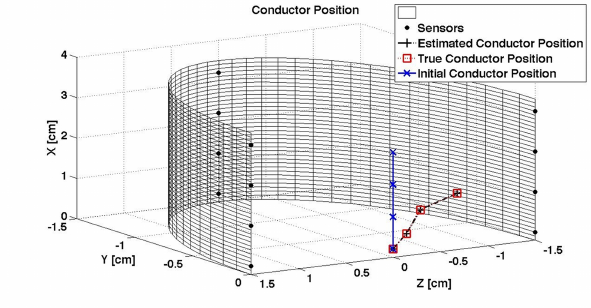}
\caption{Conductor’s position estimated by semi-cylindrical sensor panel:
the initial position is that of a straight thin-wire antenna 2.1 cm length,
perpendicular to x = 0 plane. }
\end{figure}
For the above described set up, the measured values are calculated by means of the MoM based forward solver described in section 2.
The initial set of end-points segments coordinates $s^{(0)}$ for the LM optimization algorithm is that of a straight thin-wire antenna 2.1 cm length and perpendicular to the ground plane.
From these coordinates the true position of each branch is
then estimated. Figure 3 shows the estimated position of the conductor in the case of the flat sensors panel. The picture shows that the distance between the estimated position, the black dashed line with plus markers, and the true position, the red dashed line with square markers, is close to zero.

\begin{table}[htdp]
\label{tab:flatpanel}
\caption{Relative error and number of iteration required with the flat sensors panel}
\begin{center}
\begin{tabular}{c c c}
 Current [mA] & Relative error & No. of iterations\\ \hline
 0.1					& 6.15e-11			 & 169\\
 1	 					& 6.10e-11			 & 165\\
 5 	 					& 7.88e-09			 & 87\\
 25  					& 3.78e-10			 & 28\\
 50  					& 1.53e-11			 & 8
\end{tabular}
\end{center}
\end{table}

For this configuration, Table 1 shows the $\textit{2}$-norm relative error
between true and estimated coordinates values for a given
level of current, and the number of iterations. The simulations results show that the error is close to zero, also for small values of the source current. This gives benefit for the future development of the antenna’s power system and a benefit for the patient. Figure 4 shows the estimated position of the conductor for the semi-cylindrical sensors panel. 
\begin{table}[htdp]
\label{tab:cylipanel}
\caption{Relative error and number of iteration required with the semi-cylindrical sensors panel}
\begin{center}
\begin{tabular}{c c c}
 Current [mA] & Relative error & No. of iterations\\ \hline
 0.1					& 3.05e-10			 & 25\\
 1	 					& 1.04e-09			 & 17\\
 5 	 					& 1.75e-11			 & 12\\
 25  					& 2.55e-12			 & 8\\
 50  					& 1.00e-09			 & 8\end{tabular}
\end{center}
\end{table}The picture shows as the distance between the estimated position, the black
dashed line with plus markers, and the true position, the red dashed line with square markers, is close to zero. For this scenario, the $\textit{2}$-norm relative error between true and estimated coordinates for a given level of source current, and the number of iterations are shown in Table 2. The simulations results show that the error is close to zero also in this case. 

Note that a lower relative error has been reached for the flat sensors panel:this result can be justified because with the semi-cylindrical sensors panel the measured magnetic field components contain more information about the real field distribution. Moreover
the number of iterations decreases as the source current amplitude
increases for a same sensors configuration. However, the
shape of the sensors panel influences also the convergence of
the optimization algorithm: the semi-cylindrical sensors panel
outperforms the flat sensors panel, especially for low levels of
antenna source currents. This result suggests that the number
of iterations can be reduced via a proper sensors panel design,
without compromising the precision and without increasing the antenna currents, this latter aspect is fundamental for the
invasiveness of the method.

\section{Conclusions}
In this paper, a numerical method to evaluate the shape of
micro-structures for bio-medical application with a very low
invasiveness for the human body, is proposed. A flexible thinwire
antenna radiates the VHF waves and then, by numerically
solving a typical inverse problem, the estimation of the antenna
location enables to reconstruct the micro-structure’s image.
The typical inverse problem is solved, and first simulation
results assess the validity and the robustness of the proposed
approach.\\

\end{document}